\documentclass[12pt,aasms4]{aastex}
\def\ergsec{\hbox{ergs s$^{-1}$}}
\def\ergcmsec{\hbox{ergs cm$^{-2}$ s$^{-1}$}}

\slugcomment{Accepted for publication in the {\it Astrophysical Journal Letters}}
\shorttitle{LMXRBs in M15}
\shortauthors{White \& Angelini}

\begin{document}

\title{The Discovery of a Second Luminous Low Mass X-ray Binary in
the Globular Cluster M15}
\author{Nicholas E. White, and Lorella Angelini\altaffilmark{1}}

\affil{Laboratory for High Energy Astrophysics, NASA's Goddard Space Flight Center, 
Greenbelt, MD 20771}
\altaffiltext{1}{Also Universities Space Research Association}
\email{nwhite@lheapop.gsfc.nasa.gov}
\email{angelini@davide.gsfc.nasa.gov}

\begin{abstract}

We report an observation by the {\it Chandra} X-ray Observatory of 4U2127+119, the X-ray source identified with the globular cluster M15. The {\it Chandra} observation reveals that 4U2127+119 is in fact two bright sources, separated by $2.7''$. One source is associated with AC211, the previously identified  optical counterpart to 
4U2127+119, a low mass X-ray binary (LMXB). The second source, M15-X2, is coincident with a $19^{th}$ U magnitude blue star that is $3.3"$ from the cluster core. The {\it Chandra} count rate of M15-X2 is 2.5 times higher than that of AC211.  Prior 
to the $0.5''$ imaging capability of {\it Chandra} the presence of two so closely separated bright sources would 
not have been resolved.  The optical counterpart, X-ray luminosity and
spectrum of M15-X2 are consistent with it also being an LMXB system. This is the first time that two 
LMXBs have been seen to be simultaneously active in a globular cluster.
The discovery of a second active LMXB in M15 solves a long standing puzzle where the  properties of AC211 appear consistent with it 
being dominated by an extended accretion disk corona, and yet 4U2127+119 also shows  
luminous X-ray bursts requiring that the neutron star be directly visible. 
The resolution of 4U2127+119 into two sources suggests that the X-ray bursts did not come from AC211, but rather from M15-X2. We discuss 
the implications of this discovery for understanding the origin and evolution of LMXBs in GCs as well as X-ray observations of globular clusters in nearby galaxies.

\end{abstract}



\section{Introduction}

The X-ray source 4U2127+119 associated with the globular cluster (GC) M15
was the first to be identified with an individual star within a 
GC. The identification with the $V\sim 15$ star AC211 was made 
by Auriere, le Fevre and Terzan 
(1984) using the Einstein high resolution imager (HRI) position 
(Hertz and Grindlay 1983; HG83). The
identification of AC211 with 4U2127+119 was further strengthened from a spectroscopic study 
by Charles, Jones and Naylor (1986), which showed characteristic 
signatures of a LMXB. AC211 was found to have a modulation of 8.5
hr in the optical (Ilovaisky {\it et al.} 1987), which was then subsequently
seen in the X-ray flux of 4U2127+119 (Hertz 1987; H87).  Further observations and a more detailed analysis by Ilovaisky {\it et al.} (1993; I93) revealed the true orbital period of AC211 to be 17.1 hr, twice the originally proposed value.

AC211 is optically one of the brightest known LMXBs and yet has
a relatively low X-ray luminosity of $\sim 10^{36}$ \ergsec. The high
optical to X-ray luminosity ratio suggests that a very luminous
central X-ray source is hidden behind the accretion disk, with
X-ray emission scattered into our line of sight via an 
accretion disk corona, ADC (Auriere {\it et al.} 1984, H87,
Naylor {\it et al.} 1988). 
An ADC is also required to explain
the  X-ray orbital modulation (H87, I93). 
This neat picture was put into doubt when a luminous X-ray
burst from 4U2127+119 was recorded by the Ginga satellite in 1988 (Dotani
{\it et al.} 1990). This burst was long lived ($>$ 150 s) with a precursor
event $\sim 6$~s before the longer event. The peak luminosity of the
burst was above $10^{38}$ \ergsec, with an expansion of the neutron
star photosphere (Dotani {\it et al.} 1990; van Paradijs {\it et al.} 1990),
meaning that the neutron star surface had been directly
observed. A dip in the continuum flux between the precursor and the
main burst seemed to tie this event to the X-ray source in
M15. A second X-ray burst from 4U2127+119 in September 2000, with similar properties to the first, has been reported  by Smale (2001) using RXTE. The observation of X-ray bursts from 4U2127+119 has been hard to reconcile with the idea that the
central source is hidden behind an ADC. In the {\it letter} we report the results of a {\it Chandra} High Energy Transmission Grating (HETG) observation that solves the  puzzling behavior of 4U2127+119. 


\section{Results}

4U 2127+119 was observed with the {\it Chandra} HETG grating in
conjunction with the ACIS-S CCD array on 2000 August 24, for a total
exposure of order 21,000 s. The event file was gain corrected, using
the calibration released on June 7 2001 (Caldb 2.6), screened for
bad pixels and good 
time intervals. Streaks caused by flaws in the serial readout of
the  CCDs were also removed. A CCD grade histogram shows 
a large fraction of grade seven events, indicating
considerable pile up. The zero order image from the cleaned event
file using all grades is shown in Figure 1a -- two bright sources separated by $2.7''$ can be seen. To determine the best
source positions we used data from all grades and determined the peak of the
distribution in a box of $3 \times 3$ pixels. The 
positions of the two
sources are M15-X1: R.A. (J2000) = $21^{h} 29^{m} 58{^s}.25$,
Dec (J2000) = $+12^{o} 10' 02''.9$
and M15-X2: R.A. (J2000) = $21^{h} 29^{m} 58{^s}.06$, 
Dec (J2000) = $+12^{o} 10' 02''.6$.

Accurate positions for AC211 were reported by Kulkarni {\it et al.} (1990)
based on radio measurements and by Geffert {\it et al.} (1994)
based on meridian circle measurements and the PPM catalogue. The
separation  between the radio-meridian and radio-PPM positions range
between $0.2''-0.27''$. M15-X1 is within $ 0.9''$ of the Kulkarni  position of 
R.A.(J2000)=$21^{h} 29^{m} 58{^s}.31$, Dec(J2000)=$+12^{o} 10' 02''.9$ for AC211. 
This is within the
uncertainty of the {\it Chandra} attitude solution. To obtain the best
coordinates for the second source we then corrected the image reference 
coordinates so as to give the Kulkarni position for AC211. This corresponded
to a shift of $0.94 ''$ in RA and $0.03''$ in Dec, and gives a revised
position for M15-X2 of R.A. (J2000)=$21^{h} 29^{m} 58{^s}.13$ and Dec
(J2000)=$+12^{o} 10' 02''.6$. Since the M15-X2 source shows pile-up, which could effect the centroiding, we adopted a conservative positional uncertainty 
of $0.5 ''$ diameter, based on the size of 1 pixel in the {\it Chandra} CCD. 
We designate 
the new source {\it CXO J212958.1+121002}, although continue 
to refer to it here by the more concise name M15-X2.

To search for an optical counterpart we used the {\it Hubble Space Telescope} 
(HST) images published in Guhathakurta {\it et al.} (1996). In Figure 1b we over the $0.5''$ diameter circle centered on the M15-X2 position (Figure 1b). Near the center of the error circle is a faint blue 
star. De Marchi and Paresce (1994) have cataloged the M15 stars seen in earlier HST images and in their list the faint blue star is number 590, a star with an equivalent U magnitude of 18.6.  Star 590 is $0.13 ''$ from the {\it Chandra} position centroid for M15-X2. In Figure 1c we indicate on the original color (U+B+V) image from Guhathakurta {\it et al.} (1996) the positions of AC211 and 
the blue optical counterpart to M15-X2. The new source M15-X2 is $3.4"$ from the center of M15 (the green cross in Figure 1c). 
A faint blue optical counterpart is the classic signature of a LMXB (e.g. van 
Paradijs and McClintock 1996) and it seems very likely that this is the 
counterpart to the X-ray source. 

The HEG and MEG dispersion directions and the two source positions relative to the {\it Chandra} roll angle  are such that the dispersed
grating spectra of the two sources overlap (Figure 1a). Due to the 
10 degree difference in dispersion angle between the two gratings, 
the overlap is more severe in the HEG compared to the MEG.
To extract the spectra of the two sources, histograms for the MEG and HEG
along the cross-dispersion dimension were fitted with Lorenzian models. The Lorenzian width was fixed
at the value obtained from a similar histogram of a point source.
The width of the HEG histogram is consistent with a single
source, while the MEG requires a double Lorenzian fit (see Fig 2).
We concentrated on the MEG grating spectrum because it is the best separated. 
The relative normalization 
between the two Lorenzian models for the MEG histogram is $\sim 2.5$, with M15-X2 the
brighter source. This is consistent with the fact that most of the zero order 
grade seven events, caused by pile-up, are coincident with M15-X2.
MEG spectra and light curves were accumulated by selecting regions across the
cross-dispersed direction, where the contamination was a minimum (Fig 2). The percentage of counts from each source 
included in the extracted spectra are $\sim$ 50\% for AC211 and $\sim$ 20\%
for M15-X2.

The lightcurves of the two sources exhibit quite different variability, confirming that the cross contamination is minimal (Figure 3). 
The X-ray flux of AC211 is highly variable across the observation, typical of that 
reported in the past (see e.g. Ilovaisky {\it et al.} 1987). Using the 
ephemeris given in I93 the observation began at 
orbital phase 0.6 and ended at 0.9, just prior to the predicted time 
of the partial X-ray eclipse. In contrast M15-X2 shows little variability, 
with just a very slight decline of a few percent across the 
observation. A power spectrum analysis of each source for a frequency range $4.8\times10^{-5}$ Hz to 0.196 Hz (twice the CCD read out) did not reveal any periodic signal with a peak to mean amplitude $>$10\%, neither from AC211, nor from M15-X2. There was some quasi-periodic activity on a timescale of $\sim 2.7$ hr from AC211 (which can be seen in the lightcurve) and a 700s modulation in both sources caused by a deliberate wobble in the spacecraft pointing to reduce pileup in the CCD.

The spectra of the two sources, separated by orders,
were extracted using the {\it tgextract} routine included in Ciao 2.1.
For each source the first order positive and negative MEG spectra were added together and grouped with a minimum of 20 counts per bin.
The MEG spectra of the two sources are shown in Figure 4. The X-ray spectrum
of AC211 is harder than that of M15-X2. The new source, M15-X2, can
be fit with a single power law with an energy index of $1.72 \pm 0.06$
and an absorption of $< 3.4 \times 10^{20}$ H cm$^{-2}$.
Fixing the overall absorption at the expected value to M15 equivalent 
to a hydrogen  column density of 
$6.7 \times 10^{20}$ cm$^{-2}$ still gives an acceptable fit 
(reduced $\chi^2$ of 1.14), with a power law 
index of $1.89\pm 0.05$. In contrast, AC211 is not well fit by a single
component power law or thermal bremsstrahlung model. A power law
model gives a relatively hard photon index of 1.2 with an absorption
of $6\times 10^{21}$ cm$^{-2}$, but with a reduced $\chi^2$ of 1.75. A 
partial covering model, one of several components used by Sidoli, Parmar and Oosterbroek (2000; SPO) 
to fit the broader band BeppoSAX integrated spectrum of both sources,
does provide an acceptable fit. This gives a power law index of 2.1, a
covering fraction of 0.87 with an absorption of $2\times 10^{22}$
cm$^{-2}$ and an interstellar absorption of $1.6 \times 10^{21}$ 
cm$^{-2}$. Fixing the absorption at the expected interstellar value for M15 gives
a power law index of $2.0\pm 0.1$, a covering fraction of $0.92\pm 0.01$ and an intrinsic 
absorption of $2.05\pm 0.15 \times 10^{22}$ cm$^{-2}$. This is higher than the covering fraction of 0.64 reported by SPO for the combined spectrum of both sources.
Using the expected point spread function for a single source we
corrected for the partial extraction of the two sources. The 0.5--7.0
keV flux from AC211 is $7 \times 10^{-11}$ \ergcmsec, corresponding to 
$9 \times 10^{35}$ \ergsec\ using a 
distance of 10.3 kpc to M15 (Harris 1996). For the new source, M15-X2 
the 0.5-7.0 flux is $1.1 \times 10^{-10}$ \ergcmsec, corresponding
to $1.4 \times 10^{36}$ \ergsec.

The second  burst from 4U2127+119 reported by Smale (2001), occurred one month after the {\it Chandra} observation -- indicating the burst source to still be active. The RXTE all sky monitor 
(ASM) light curve of M15 does not show any transient outburst or unusual behavior around 
the time of the {\it Chandra} and RXTE observations. 
The 5yr long ASM light curve is relatively steady, suggesting that the source is long lived and not highly variable. There are short outbursts in the ASM every 
$\sim 365$ days, but they occur when the source passes close 
to the sun and are probably caused by sun glint on the detector 
collimator upsetting the solutions. It is interesting 
to note that the original $\pm 1''$ position from the Einstein HRI (HG83) 
lies closer to that of M15-X2, than AC211, and that the search 
for the optical counterpart by Auriere {\it at al.} (1984) used a circle 
of $3.3''$, the standard HRI 90\% confidence error circle.
The ROSAT HRI archival images from the mid-1990s appear somewhat 
elongated in the East-West direction --
suggesting the M15-X2 source may have been present.

\section{Discussion}

This is the first time that two LMXBs have been seen to be
simultaneously active in a globular cluster associated with our Galaxy. The
separation of $ \sim 2.7''$ is less than the resolution of previous X-ray
telescopes. It is only with the superb $0.5''$ quality imaging
of the {\it Chandra} X-ray Observatory that 4U2127+119 can be resolved into two sources. 
The new 
X-ray source M15-X2 is 2.5 times brighter than AC211 and associated with a faint 
$19^{th}$ U magnitude blue star, characteristic of a LMXB. It is 3.4 arc from the center of M15 (Figure 1c). The discovery of a  second active LMXB in M15
provides a simple explanation for the apparently schizophrenic 
properties of 4U2127+117 -- M15-X2 is
the source of the X-ray bursts, not the ADC dominated AC211. The 
presence of a second source also resolves why the spectrum of 
4U2127+119 is unusually 
complex (SPO). The X-ray continuum spectrum of M15-X2 is a power law with a photon index of $\sim 1.9$ and a luminosity of $\sim 10^{36}$ \ergsec,  
both typical of LMXBs that show X-ray bursts. RXTE and archival data suggests that the 2nd source may have been present over the entire past 20 yr. This confirms earlier suspicions that two sources could explain the conflicting properties of 4U2127+117 (Grindlay 1992, 1993; Charles, van Zyl \& Clarkson 2001). The ADC  model to explain the X-ray and optical properties of AC211 is now self consistent. The X-ray spectrum of AC211 alone is harder and resembles that of the classic ADC source 4U1822-371 (White {\it et al.} 1981; White, Kallman and Angelini 
1997). The average X-ray luminosity of AC211 is really one third lower than previously 
thought. This both increases the amplitude of the orbital modulation 
and reduces the luminosity -- strengthening the analogy with 4U1822-371 (White and Holt 1982). 

The ratio of LMXBs to stellar mass is more 
than two orders of magnitudes higher for globular clusters than
it is for the rest of the Galaxy (Clark 1975). This overabundance of
LMXBs led Fabian, Pringle \& Rees (1975) to propose that the LMXBs in
globular clusters are formed via tidal capture of neutron stars 
in close encounters with main-sequence or giant stars, a
mechanism that operates efficiently in the high stellar density
found in globular clusters.  Hut, Murphy and Verbunt (1991; HMV) discuss the
probability of finding one or more LMXBs in any particular globular cluster. 
This depends strongly on how many neutron stars stay in the
cluster after they are born and the lifetime of the LMXBs. In general
these calculations suggest that more than one LMXB should be found 
in a globular cluster. The fact that so few LMXBs in globular clusters are observed has required either 
short lifetimes (HMV) or larger fractions of neutron
stars ejected from the globular cluster (Verbunt and Hut 1987).  The large number of millisecond radio pulsars found in globular clusters, thought to be the remains of a LMXB
systems, also points to many LMXBs having been active in the past
(see HMV). {\it Chandra} observations of other GCs have revealed 
faint sources some of which may be a transient LMXB in quiescence, suggesting that there are  more LMXB in a single GC ({\it e.g.} Grindlay {\it at al.} 2001). This discovery of two active LMXBs in M15 moves the observations in the right direction with respect to the theory and
number of radio pulsars, especially given the small number of globular cluster systems  in our Galaxy with active LMXBs. 

{\it Chandra} X-ray observations of nearby galaxies have identified many
point X-ray sources with globular clusters (Sarazin, Irwin, \& Bregman 2001,
Angelini, Loewenstein, \& Mushotzky 1991). 
Many of these GC sources have a luminosity above the Eddington 
limit for accretion onto a neutron star. Angelini {\it et al.} (1991) have suggested that some of these high luminosity systems may be due to multiple LMXBs being active in some
GC systems. The discovery of two LMXBs active at the
same time in M15 adds weight to that argument.

Acknowledgements. We recognize the critical contribution of Cynthia 
Hess, the original principle investigator for this observation. We 
thank Ian George and Roy Kilgard for help with the grating analysis, Brian Yanny 
for providing an HST image of M15, Phil Charles for  
discussions on the optical counterpart, Robin Corbet for advice on the RXTE ASM and Josh Grindlay, Peter Edmonds and Sergio Ilovaisky for comments on the manuscript. This research made used of data extracted from the HEASARC, MAST and CDS.


\small

\begin{figure}
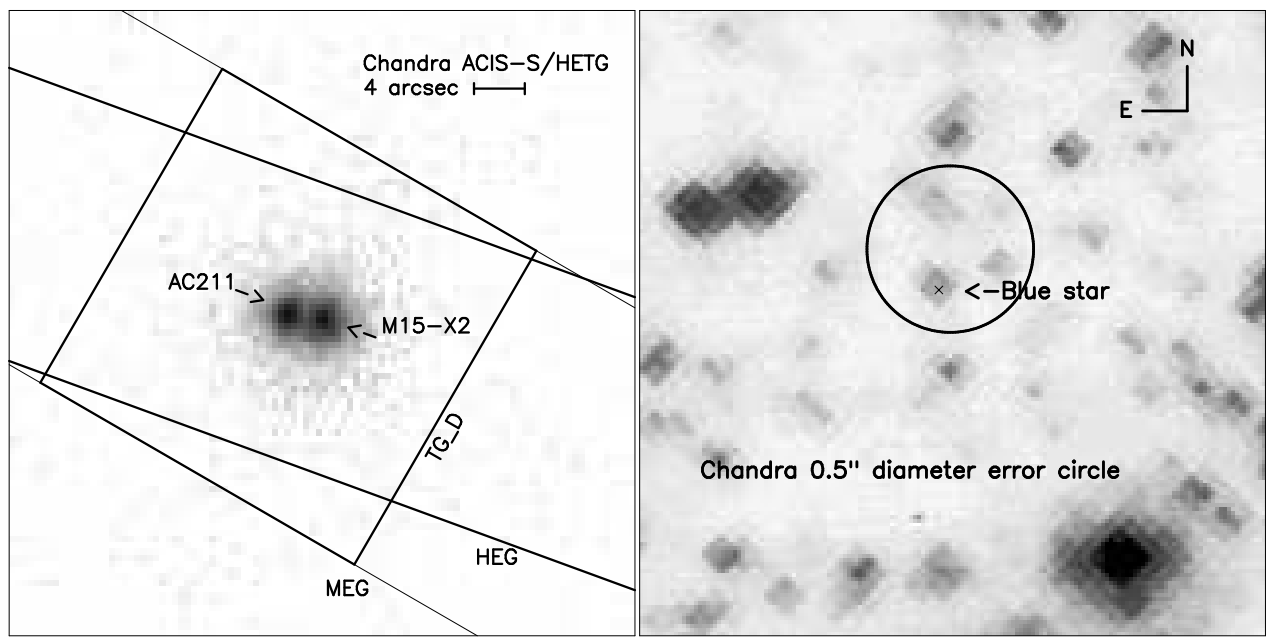

\figurenum{1a \& 1b}
\centerline{
\includegraphics[scale=0.6,angle=-90]{fig1a_ximg.eps}\includegraphics 
[scale=0.6,angle=-90]{fig1b_oimg.eps}}
\caption{{\it a) Left :} The ACIS-S/HETG {\it Chandra} zero order image 
of M15 showing the two sources. The tick lines across 
the image mark the dispersion directions for the MEG and the HEG, which are 
separated by 10 degrees. The dispersion direction
for the MEG and HEG with respect to M15-X2 is offset by $\sim 5$~pixels
in the X direction and $\sim 1$~pixel in the Y direction.
The perpendicular line marked with TG\_D (shown only for the MEG)
indicates the cross-dispersion direction. The MEG count histogram 
(shown in Figure 2) is accumulated with respect to that dimension. 
{\it b) Right :} An enlargement of the 
U+B+V HST image taken from Guhathakurta et al. (1996) centered on the 
{\it Chandra} position of M15-X2. 
The {\it blue star} corresponds to star number 590 in 
De Marchi \& Paresce (1994).}
\end{figure}

\begin{figure}
\figurenum{1c}
\centerline{\includegraphics[scale=1.0,angle=0]{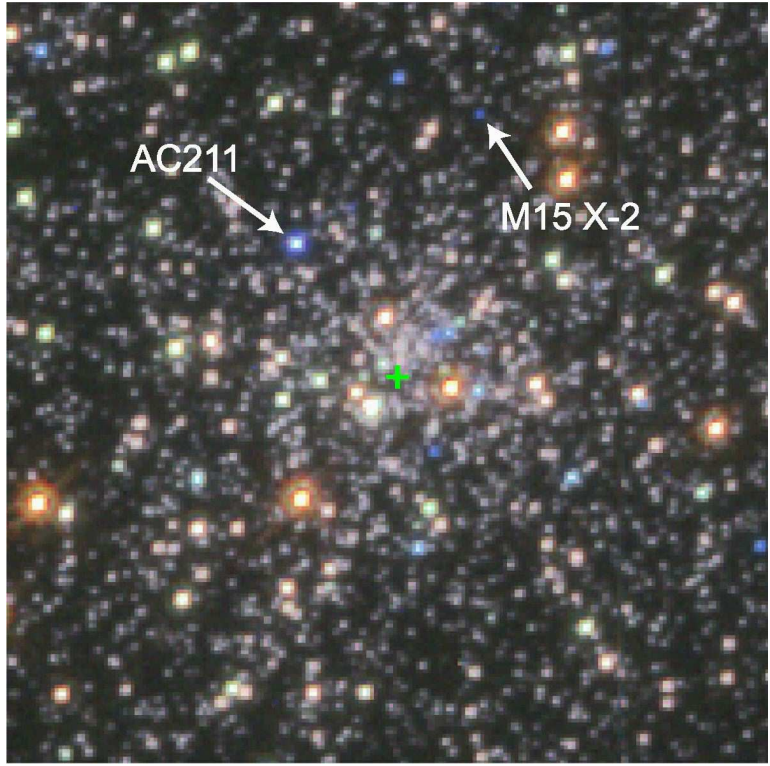}}
\caption{A true color (U+B+V) Hubble Space Telescope image of the central 9"x9" region of M15 reproduced from Guhathakurta et al. (1996). The positions of AC211 and the $\sim 19^{th}$ U magnitude blue star identified with M15-X2 are indicated. The green cross shows the center of M15 (Guhathakurta et al. 1996). Note the orientation is different from Fig 1b.}
\end{figure}

\begin{figure}
\figurenum{2}
\centerline{\includegraphics[scale=0.6, angle=0]{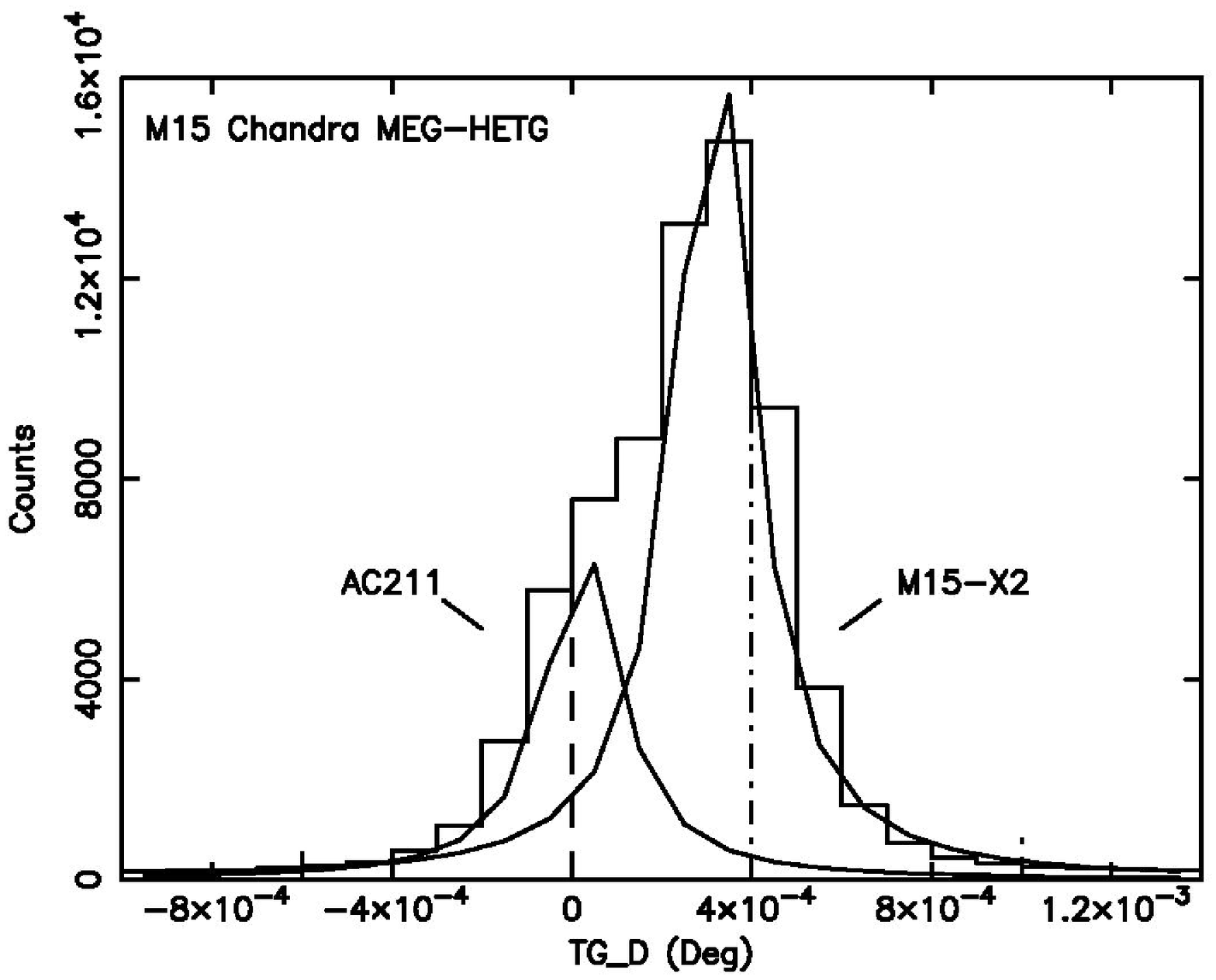}}
\caption{Histogram of the MEG cross-dispersion direction (TG\_D)
with respect to AC211, shown with the best Lorenzian fits to model the
two sources. The dash and dash-dotted lines mark the regions selected 
to extract the spectra for AC211 and M15-X2. These are taken from 
the tail of the distribution to minimize contamination. 
The X-axis boundary of the M15-X2 region are shifted relative to the 
AC211 position to give an overall view from where the spectra where
taken with  respect to TG\_D.}
\end{figure}

\begin{figure}
\figurenum{3}
\centerline{\includegraphics[scale=0.5,angle=-90]{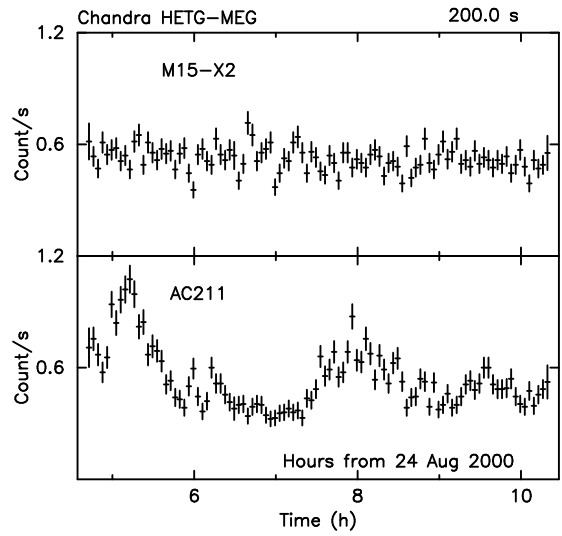}}
\caption{The 0.3-10 keV light curves of the two sources using the same extraction
regions used for the spectra. These are not corrected for the extraction region used and 
the count rates must be scaled up by $\sim 2$ and $\sim 5.5$ for AC211 and M15-X2, respectively. The {\it Chandra} observation covers from phase 0.6 to 0.9
of the AC211 orbit, using the ephemeris given by I93.}
\end{figure}

\begin{figure}

\figurenum{4}
\centerline{\includegraphics[scale=0.5,angle=-90]{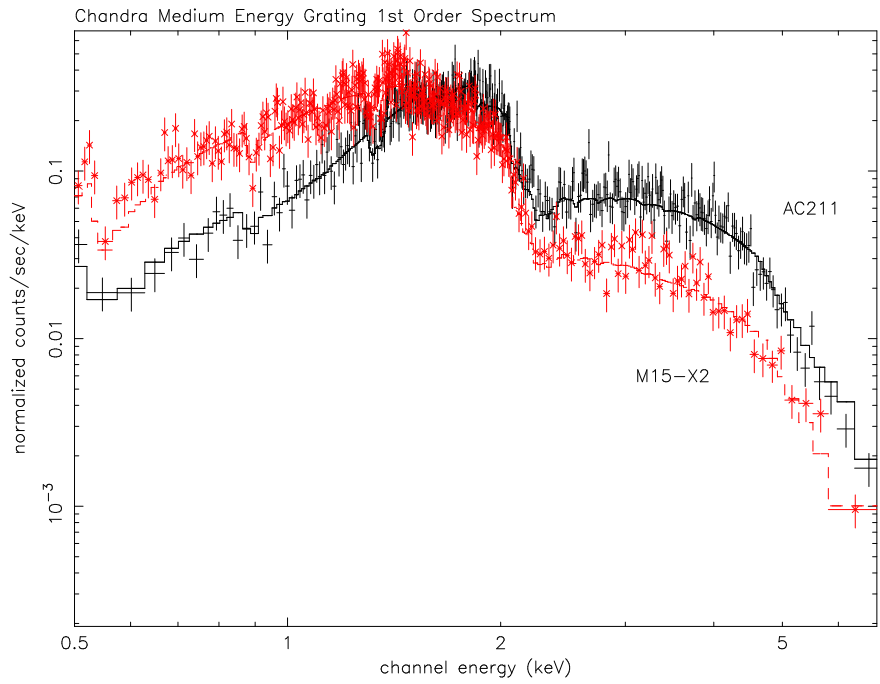}}
\caption{The MEG spectra for M15-X2 (dashed) and AC211 (solid) with 
the best fit models shown as histograms. The overall normalization is not corrected
for the extraction region used. The correction factors for AC211 and M15-X2 are $\sim 2$ and $\sim 5.5$, respectively}
\end{figure}

\end{document}